\documentclass[prl,aps,groupedaddress,floatfix,twocolumn,superscriptaddress,showpacs]{revtex4}
\usepackage{amsmath,amssymb,epsfig,bm,pifont}
\usepackage[linkcolor=black,citecolor=black,urlcolor=blue,colorlinks=true,
linktocpage=true]{hyperref}
\usepackage{graphicx}
\usepackage{epstopdf}
\usepackage{epsfig}
\usepackage{bm}
\usepackage{tabularx}

\newcommand{\beq}{\begin{equation}}
\newcommand{\eeq}{\end{equation}}
\newcommand{\bea}{\begin{eqnarray}}
\newcommand{\eea}{\end{eqnarray}}
\newcommand{\tr}{\mathrm{tr}}

\begin{document}

\title{Entanglement, noise, and the cumulant expansion}

\author{Joaqu\'{\i}n E. Drut}
\email{drut@email.unc.edu}
\affiliation{Department of Physics and Astronomy, University of North Carolina,
Chapel Hill, North Carolina, 27599-3255, USA}

\author{William J. Porter}
\email{wjporter@live.unc.edu}
\affiliation{Department of Physics and Astronomy, University of North Carolina,
Chapel Hill, North Carolina, 27599-3255, USA}

\begin{abstract}
We put forward a simpler and improved variation of a recently proposed method to 
overcome the signal-to-noise problem found in Monte Carlo calculations of
the entanglement entropy of interacting fermions.
The present method takes advantage of the approximate lognormal distributions that
characterize the signal-to-noise properties of other approaches. In addition,
we show that a simple rewriting of the formalism allows circumvention of the inversion of the restricted
one-body density matrix in the calculation of the $n$-th R\'enyi entanglement entropy for $n>2$.
We test our technique by implementing it in combination with the hybrid Monte Carlo algorithm and
calculating the $n=2,3,4, \dots, 10$ R\'enyi entropies of the 1D attractive Hubbard model.
We use that data to extrapolate to the von Neumann ($n=1$) and $n\to\infty$ cases.
\end{abstract}

\date{\today}
\maketitle

\section{Introduction} 

Recently~\cite{HMCEE}, we proposed an algorithm to compute the R\'enyi entanglement entropy $S_n$
of interacting fermions. Many algorithms have been proposed to this effect in the last few 
years~\cite{Buividovich, Melko, Humeniuk, McMinis, Broecker, WangTroyer, Luitz}.
Our proposal, based on the free-fermion decomposition approach of Ref.~\cite{Grover}, overcomes the 
signal-to-noise problem present in that approach and is compatible with the hybrid Monte Carlo (HMC) 
method~\cite{HMC} widely used in the context of
lattice quantum chromodynamics. The core idea of our method is that, by differentiating with respect to an 
auxiliary parameter $\lambda$, one may carry out a Monte Carlo (MC) calculation of $d S_n/d\lambda$ with a probability 
measure that includes entanglement properties explicitly. [This was not the case in the approach of Ref.~\cite{Grover}, 
where the probability measure factored across auxiliary field replicas; we identified this as the cause of the
signal-to-noise problem (see below)]. Once the MC calculation is done, integration 
with respect to $\lambda$ returns the desired entanglement entropy relative to that of a noninteracting 
system (which is easily computed separately).

In this work, we describe and implement a variation on that Monte Carlo algorithm which, 
while sharing the properties and core idea mentioned above, differs from it in two important ways; the
new method, in fact, is different enough that we advocate its use over our original proposal.
First, the new method takes advantage of the approximate lognormal shape
of the underlying statistical distributions of the fermion determinants, which we already noted in
Ref.~\cite{HMCEE} and which we explain in detail below.
Second, and more importantly, the present method is simpler than our original proposal: whereas in the latter the 
parameter $\lambda$ multiplied the coupling constant $g$ (thus generating a rather involved set of terms upon 
differentiation of the fermion determinant), here $\lambda$ is coupled to the number of fermion species $N_f^{}$. 
As we show below, this choice not only simplifies the implementation, but also exposes the central role of the logarithm
of the fermion determinant in our calculation of $S_n$, and thus brings to bear the approximate lognormality
property mentioned above.

Below, we present the basic formalism, review the evidence for approximate lognormal distributions, and  
explain our method. Besides the points mentioned above, in our calculations we have found the present method 
to be more numerically stable than its predecessor. We explain this in detail in our Results section.

In addition to the new method, we show that it is possible to rewrite part of the formalism in order to bypass the
calculation of inverses of the restricted density matrix (see e.g.~\cite{Broecker,WangTroyer,HMCEE})
in the determination of R\'enyi entropies of order $n>2$. 
To test our method, we computed the $n=2$ R\'enyi entropy of the 1D attractive Hubbard model
using the previous as well as the new formalism, and checked that we obtained identical results. 
Going beyond the $n=2$ case, we present results for the $n=2,3,4,\dots, 10 $ R\'enyi entropies and 
find that higher R\'enyi entropies display lower statistical uncertainty in MC calculations.

\section{Basic formalism}

As in our previous work, we set the stage by briefly presenting the formalism of Ref.~\cite{Grover}.
The $n$-th R\'enyi entropy $S^{}_{n}$ of a sub-system $A$ of a
given system is
\beq
\label{Eq:SnDef}
S^{}_{n}= \frac{1}{1-n} \ln \tr (\hat \rho^n_A),
\eeq
where $\hat  \rho^{}_A$ is the reduced density matrix of sub-system $A$. 
For a system with density matrix $\hat \rho$, 
the reduced density matrix is defined via a partial trace over the Hilbert space 
corresponding to the complement $\bar{A}$ of our sub-system:
\beq
\label{Eq:RedDM}
\hat{\rho}^{}_{A} = \tr_{\bar{A}}\hat{\rho}.
\eeq

An auxiliary-field path-integral form for
$\hat \rho^{}_A$, from which $S^{}_{n}$ can be computed using MC methods 
for a wide variety of systems, was presented in Ref.~\cite{Grover}, which we 
briefly review next.

As is well known from conventional many-body formalism, the full density matrix $\hat \rho$ can be written as a path integral
by means of a Hubbard-Stratonovich auxiliary-field transformation:
\beq
\label{Eq:RhoSigma}
\hat \rho = \frac{e^{-\beta \hat H}}{\mathcal Z} = \int \mathcal D \sigma^{} P[\sigma]\, \hat \rho[\sigma],
\eeq
for some normalized probability measure $P[\sigma]$ determined by the details of the underlying Hamiltonian 
(for more detail, see below and also Ref.~\cite{MCReviews}). Here, $\mathcal Z$ is the partition function, and $\hat \rho[\sigma]$ is 
the density matrix of noninteracting particles in the external auxiliary field $\sigma$. One of the main contributions
of Ref.~\cite{Grover} was to show that the above decomposition determines not only the full density matrix but also the restricted 
one. Indeed, Ref.~\cite{Grover} shows that
\beq
\label{Eq:rhoA}
\hat{\rho}^{}_{A} = \int \mathcal D\sigma^{}P[\sigma]\,\hat{\rho}^{}_{A}[\sigma],
\eeq
where $P[\sigma]$ is the same probability used in Eq.~(\ref{Eq:RhoSigma}),
\beq
\label{Eq:rhoAcdaggerc}
\hat{\rho}^{}_{A}[\sigma] = C_{A}[\sigma]\; \exp\left(-\sum_{i,j} \hat c^{\dagger}_i [\ln(G^{-1}_{A}[\sigma^{}]-\openone)]_{ij}^{} \hat c^{}_j\right),
\eeq
and
\beq
C_{A}[\sigma] = \det(\openone - G^{}_{A}[\sigma^{}]).
\eeq

Here, $G^{}_{A}[\sigma^{}]$ is the restricted Green's function of the noninteracting system in the external field $\sigma$
(see below), and $\hat c^{\dagger}$, $\hat c^{}$ are the fermion creation and annihilation operators. The sums in the
exponent of Eq.~(\ref{Eq:rhoAcdaggerc}) go over those points in the system that belong to the subsystem $A$.

Using the above formalism for the case of $2N$-component fermions, the entanglement entropy (c.f. Eq.~\ref{Eq:SnDef}) 
takes the form
\bea
\label{Eq:SnMC}
\exp\text{\big(}(1-n)S^{}_{n} \text{\big)} =
\int \mathcal D {\{\sigma\}}^{} P[\{\sigma \}]\,Q[\{\sigma \}],
\eea
where the field integration measure, given by 
\beq
\mathcal D {\{\sigma\}}^{}  = 
\prod_{k=1}^{n} \frac{\mathcal D {\sigma^{}_k}}{\mathcal Z},
\eeq
is over the $n$ ``replicas'' $\sigma^{}_k$ of the 
Hubbard-Stratonovich field (which result from taking the $n$-th power of the path
integral representation of $\hat{\rho}^{}_{A}$ shown above), and
the normalization
\beq
\mathcal Z = 
\int \mathcal D {\sigma}^{} \prod_{m=1}^{2N} {{\det}\,U^{}_{m}[\sigma]}
\eeq
was included in the measure. 
It is worth noting that, by separating a factor of $\mathcal Z^n$ in the denominator of
Eq.~(\ref{Eq:SnMC}), an explicit form can be identified in the numerator as in the replica
trick~\cite{CalabreseCardy}, which corresponds to a partition function for $n$ copies of the system, 
``glued'' together in the region $A$.

The naive probability measure, namely
\beq
\label{Eq:PP}
P[\{\sigma \}] = \prod_{k=1}^{n} \prod_{m=1}^{2N} {{\det}\,U^{}_{m}[\sigma^{}_k]},
\eeq
factorizes across replicas, which makes it insensitive to entanglement. 
This factorization is the main reason why using $P[\{\sigma \}]$ as a MC probability 
leads to signal-to-noise issues (see Ref.~\cite{Grover}). 
In Eq.~(\ref{Eq:PP}), $U_m[\sigma]$ encodes the dynamics of the $m$-th
fermion component, including the kinetic energy and the form of the interaction after a
Hubbard-Stratonovich transformation. That matrix also encodes the form of the trial state 
$|\Psi \rangle$ in ground-state approaches (see e.g. Ref.~\cite{MCReviews}), which we use here; 
we have taken $|\Psi \rangle$ to be a Slater determinant. In finite-temperature approaches, $U_m[\sigma]$ 
is obtained by evolving a complete set of single-particle states in imaginary time.

The quantity that contains the pivotal contributions to entanglement is 
\beq
\label{Eq:QQ}
Q[\{\sigma \}] = \prod_{m=1}^{2N} {{\det}\,M^{}_{m}[\{\sigma \}]},
\eeq
which we refer to below as the ``entanglement determinant,'' and where
\bea
\label{MatrixM}
M_{m}[\{\sigma\}] &\equiv& \prod_{k=1}^{n} \left(\openone - G^{}_{A,m}[\sigma^{}_k]\right)\times \nonumber \\ 
&&\left[\openone + \prod_{k=1}^{n}\frac{G^{}_{A,m}[\sigma^{}_k]}{\openone - G^{}_{A,m}[\sigma^{}_k]} \right].
\eea
The product $Q[\{\sigma \}]$ played the role of an observable in Ref.~\cite{Grover},
which is a natural interpretation given Eq.~(\ref{Eq:SnMC}). However, we will interpret this
differently below.  Other than the field replicas, the new ingredient in the determination of $S^{}_n$ is
the {\it restricted Green's function} $G^{}_{A,m}[\sigma^{}_k]$. This is the same as the noninteracting 
one-body density matrix $G(x,x')$ of the $m$-th fermion component in the background field $\sigma^{}_k$,
but the arguments $x,x'$ are restricted to the region $A$ (see Ref.~\cite{Grover} and also 
Ref.~\cite{Peschel}, where expressions were originally derived for the 
reduced density matrix of noninteracting systems, based on reduced Green's functions). 

\section{Avoiding inversion of the reduced Green's function for $n>2$}

As noted in Ref.~\cite{Assaad}, for $n\!=\!2$, no inversion of $\openone - G^{}_{A,m}[\sigma^{}_k]$ is actually
required in the calculation of the entanglement determinant $Q[\{\sigma \}]$, as the equations clearly simplify in that case.
However, for higher $n$ it is not obvious how to avoid such an inversion. Here, however, 
we show that this calculation can indeed be accomplished {\it without} inversion.
We begin by noting that
\beq
\label{detM}
{{\det}\,M^{}_{m}[\{\sigma \}]} = {{\det}\,L^{}_{m}[\{\sigma \}]} {{\det}\,K^{}_{m}[\{\sigma \}]},
\eeq
where
$L_m[\{\sigma\}]$ is a block diagonal matrix (one block per replica $k$):
\beq
\label{MatrixL}
L_m[\{\sigma\}] \equiv \text{diag}\left[\openone-G^{}_{A,m}[\sigma^{}_k]\right],
\eeq
and
\beq
\label{MatrixK}
K_m[\{\sigma\}] 
\equiv 
\left( \begin{array}{ccccccc}
\openone & 0 & 0 & \dots & 0 & \!\!\!-R[\sigma_n]\\
R[\sigma_1] & \openone & 0 & \dots & \vdots & \!\!\! 0 \\
0 & R[\sigma_2] & \openone & 0 & 0 & \!\!\! 0 \\
\vdots & \ddots & \ddots & \ddots & \openone & \!\!\! \vdots \\
0 & \dots & \dots & 0 & R[\sigma_{n-1}]  & \!\!\! \openone
\end{array} \right),
\eeq
with
\beq
R[\sigma_k] = \frac{G^{}_{A,m}[\sigma^{}_k]}{G^{}_{A,m}[\sigma^{}_k] - \openone}.
\eeq
The equivalence of the determinants in Eq.~(\ref{detM}) can be shown in a straightforward fashion:
the $L_m[\{\sigma\}]$ factor is easily understood, as that matrix is block diagonal and therefore its
determinant reproduces the first r.h.s. factor in the first line of Eq.~(\ref{MatrixM}); the remaining factor
relies on the identity
\beq
\label{IdentityU}
\det \!\left( \begin{array}{ccccccc}
\!\!\!\openone\!\!\! & \!\!\!0\!\!\! & \!\!\!0\!\!\! & \!\dots\!\!\! & \!\!\!0\!\!\! & \!\!\!H_k\\
\!\!\!-H_1\!\!\! & \!\!\!\openone\!\!\! & \!\!\!0\!\!\! & \!\dots\!\!\! & \!\!\!\vdots\!\!\! & \!\!\! 0 \\
\!\!\!0\!\!\! & \!\!\!-H_2\! & \!\!\!\openone\!\!\! & \!0\!\!\! & \!\!\!0\!\!\! & \!\!\! 0 \\
\!\!\!\vdots\!\!\! & \!\!\!\ddots\!\!\! & \!\!\!\ddots\!\!\! & \!\ddots\!\!\! & \!\!\!\openone\!\!\! & \!\!\! \vdots \\
\!\!\!0\!\!\! & \!\!\!\dots\!\!\! & \!\!\!\dots\!\!\! & \!0\!\!\! & \!-H_{k-1}\!\!\!  & \!\!\! \openone
\end{array} \right)
\!=\!
\det\left( \openone + H_1 H_2 \dots H_k \right)
,
\eeq
which is valid for arbitrary block matrices $H_j$, is a standard result often used in many-body physics 
(especially when implementing a Hubbard-Stratonovich transformation),
and can be shown using so-called elementary operations on rows and columns.

Within the determinant of Eq.~(\ref{detM}), we may of course multiply $K_m[\sigma]$ and $L_m[\sigma]$:
\beq
\label{MatrixT}
T_m[\{\sigma\}] \equiv K_m[\{\sigma\}] L_m[\{\sigma\}] = \openone - \mathcal G_m[\{\sigma\}] B,
\eeq
where $\mathcal G_m[\{\sigma\}]$ is a block diagonal matrix defined by
\beq
\mathcal G_m[\{\sigma\}] = \text{diag}\left[{G^{}_{A,m}[\sigma^{}_{n}] }\right],
\eeq
and
\beq
\label{MatrixB}
B \equiv 
\left( \begin{array}{ccccccc}
\openone& 0 & 0 & \dots & -\openone\\
\openone & \openone & 0 & \dots & 0 \\
0 & \openone & \openone  & \dots & 0 \\
\vdots & \ddots & \ddots & \ddots & \vdots \\
0 & \dots & 0 & \openone & \openone
\end{array} \right).
\eeq
Equation~(\ref{MatrixT}) shows our claim, as we may use $T_m[\{\sigma\}]$ in our calculations 
instead of $M_m[\{\sigma\}]$, and the former contains no inverses of $\openone - G^{}_{A,m}$.

Summarizing, a class of approaches to calculating $S_n$ for $n>2$, based on the
Hubbard-Stratonovich representation of $\hat \rho_A$ (also known as free-fermion decomposition), 
requires computing $M_m[\{\sigma\}]$, which in turn requires inverting $\openone - G^{}_{A,m}$ per
Eq.~(\ref{MatrixM}). By arriving at Eq.~(\ref{MatrixT}), and given that
\beq
\det T_m[\{\sigma\}] = \det M_m [\{\sigma\}],
\eeq
[Eq.~(\ref{detM}) and beyond] we have shown that no inversions are actually required, as $T_m[\{\sigma\}]$ contains 
no inverses. While this is a desirable feature from a numerical point of view, it should be mentioned that, from a 
computational-cost point of view, the price of not inverting $\openone - G^{}_{A,m}$ 
reappears in the fact that $T_m$, though sparse, scales linearly with $n$ in size.

For the remaining of this work, calculations carried out at $n=2$ use the $M$ approach,
which is based on Eq.~(\ref{MatrixM}) and the `proposed method' described below. 
We reproduced those results by switching to the $T$ approach, which uses Eq.~(\ref{MatrixT}) (as well
as the method described below), and then proceeded to higher $n$ with the latter.

\section{A statistical observation: lognormal distribution of the entanglement determinant}

\begin{figure}[t]
\includegraphics[width=1.0\columnwidth]{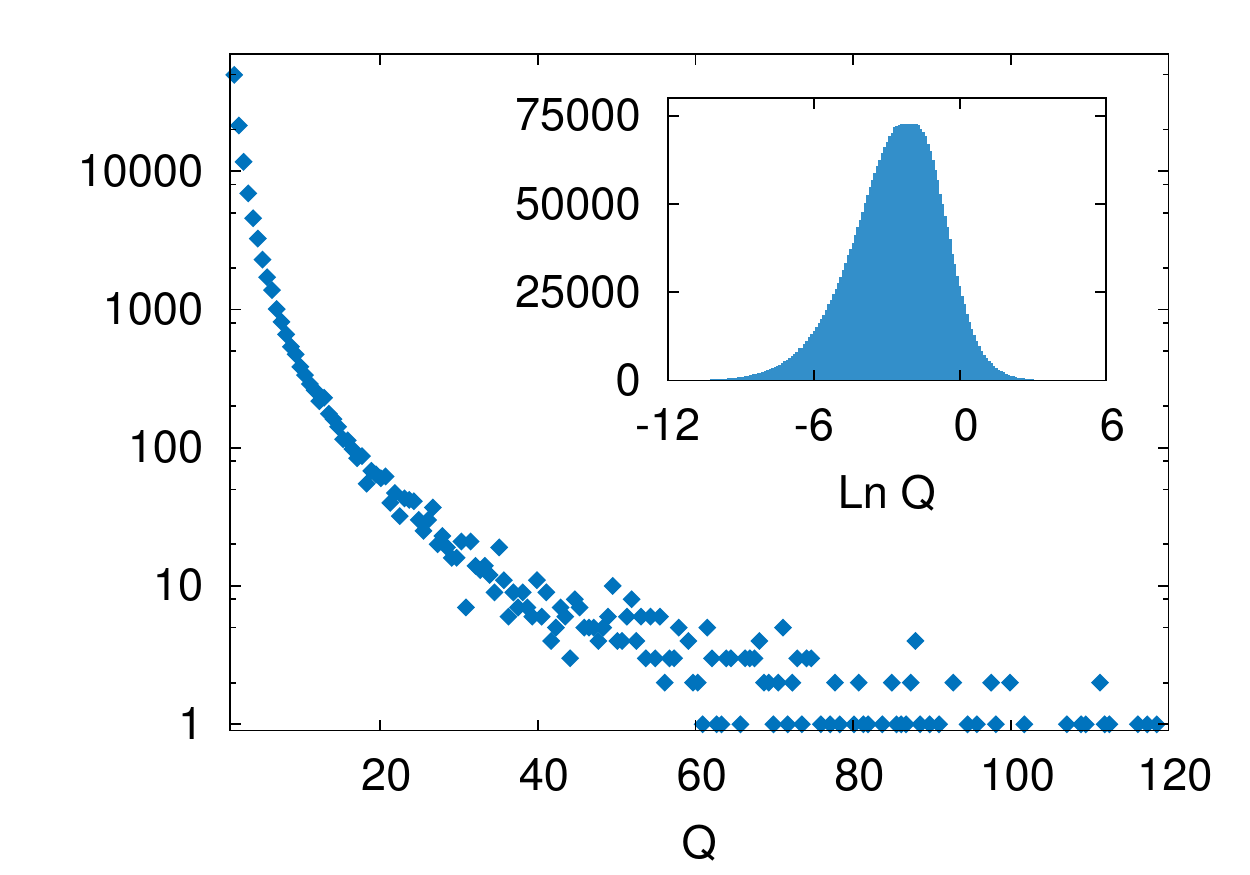}
\caption{\label{Fig:NaiveHistogram}(color online) 
Distribution of the observable $Q[\{\sigma\}]$ of the naive free-fermion decomposition method, i.e. using Eq.~(\ref{Eq:SnMC}), for a ten-site Hubbard model described by Eq.~(\ref{Eq:HubbardHam}),
at attractive coupling $U/t=2.0$ and for a subsystem of size $L^{}_A/L = 0.8$. Here, $Q[\{\sigma\}]$ is a non-negative quantity.
The long tail in the main plot (note logarithmic scale in vertical axis) is approximately a log-normal 
distribution [i.e. $\ln Q[\{\sigma\}]$ is roughly a normal distribution (see inset)].
}
\end{figure}

In Ref.~\cite{HMCEE}, we presented examples of the approximate log-normal distributions obeyed by $Q[\{\sigma \}]$
when sampled according to $P[\{\sigma \}]$. One such example is reproduced here for reference in Fig.~\ref{Fig:NaiveHistogram}.
The fact that such distributions are approximately log-normal, at least visually, suggests that one may use the 
cumulant expansion to determine $S_n^{}$. Indeed, in general, 
\bea
(1-n)S^{}_{n} &=& \ln \int \mathcal D {\{\sigma\}}^{} P[\{\sigma \}]\,Q[\{\sigma \}] \nonumber \\
&=& \sum_{m=1}^{\infty} \frac{\kappa_m^{}[\ln Q]}{m!},
\eea
where $\kappa_m^{}[\ln Q]$ is the $m$-th cumulant of $\ln Q$, and the first two nonzero cumulants are given by
\beq
\kappa_1^{}[X] = \langle X\rangle
\eeq
and
\beq
\kappa_2^{}[X] = \langle X^2\rangle - \langle X\rangle^2
\eeq
for a functional $X[\{\sigma\}]$, and where the expectation value $\langle\, \cdot\, \rangle$ here taken with respect to the produce measure $P[\{\sigma\}]$. If the distribution of $\ln Q$ were truly gaussian, the above
series would terminate after the first two terms, which would provide us with an efficient way to bypass signal-to-noise issues
in the determination of $S^{}_{n}$ with stochastic methods~\cite{NoiseSignProblemStatistics}. Unfortunately,
the distribution is not exactly gaussian. Moreover, the cumulants beyond $m=2$ are often extremely sensitive to the details
of the distribution (i.e. they can fluctuate wildly), they are hard to determine stochastically (the signal-to-noise problem re-emerges),
and there is no easy way (that we know of) to obtain analytic insight into the large-$m$ behavior of $\kappa^{}_m$.
However, this approximate log-normality does provide a path forward, as it indicates that we may still evaluate 
$\langle \ln Q \rangle$ with good precision with MC methods. As we will see in the next sections, this is enough to determine 
$S^{}_{n}$ if we are willing to pay the price of a one-dimensional integration on a compact domain.

Although (approximate) lognormality in the entanglement determinant seems very difficult to prove analytically in the present case,
evidence of its appearance has been found in systems as different as ultracold atoms and relativistic gauge 
theories~\cite{NoiseSignProblemStatistics,LogNormalDeGrand}. The underlying reason for this distribution appears
to be connected to a similarity between the motion of electrons in disordered media and lattice fermions in the
external auxiliary (gauge) field in MC calculations.

\section{Proposed method}

Starting from the right-hand side of Eq.~(\ref{Eq:SnMC}), we introduce an auxiliary 
parameter $0\le\lambda \le 1$ and define a function $\Gamma(\lambda;g)$ via 
\beq
\label{Eq:GammaDef}
\Gamma(\lambda;g) \equiv \int \mathcal D {\{\sigma\}}^{} P[\{\sigma \}]\;
Q^{\lambda}[\{\sigma \}].
\eeq
At $\lambda=0$,
\beq
\ln \Gamma(0;g)=0,
\eeq
while for $\lambda = 1$, $\Gamma(\lambda;g)$ yields
the entanglement entropy:
\beq
\frac{1}{1-n}\ln \Gamma(1;g)=S_{n}^{}.
\eeq

Using Eq.~(\ref{Eq:GammaDef}), 
\beq
\label{Eq:dlnGammadlambda}
\frac{\partial \ln \Gamma}{\partial \lambda}=
\int \mathcal D {\{\sigma\}}^{} \tilde P[\{\sigma \};\lambda]\; \ln Q[\{\sigma \}]
\eeq
where
\beq
\label{Eq:Ptilde}
\tilde{P}[\{\sigma \};\lambda]=\frac{1}{\Gamma(\lambda;g)}P[\{\sigma \}]\;Q^{\lambda}[\{\sigma \}].
\eeq
In the presence of an even number of flavors $2N$ and attractive interactions, 
$P[\{\sigma \}]$ and $Q[\{\sigma\}]$ are real and non-negative for all $\sigma$, such 
that there is no sign problem and $\tilde P[\{\sigma \};\lambda]$ above is a well-defined, normalized probability measure.

As in our previously proposed method, we can then calculate $S^{}_n$ by taking the $\lambda=0$ point
as a reference and computing $S_n$ using
\beq
\label{Eq:SnMCFinal}
S_{n}^{} = \frac{1}{1-n}
\int_{0}^{1}d\lambda\;\langle\ln{Q}[\{\sigma \}]\rangle_\lambda^{},
\eeq
where
\beq
\langle X\rangle_\lambda^{}=\int \mathcal D {\{\sigma\}}^{} \tilde{P}[\{\sigma\};\lambda]\; X[\{\sigma \}].
\eeq
We thus obtain an integral form of the interacting R\'enyi entropy
that can be computed using any MC method (see e.g.~\cite{MCReviews}), in particular HMC~\cite{HMC}.

As in our previous work, we note that the above expectation values are determined with respect 
to the probability measure $\tilde{P}[\{\sigma \};\lambda]\;$, which communicates correlations 
responsible for entanglement. In contrast to the canonical MC probability $P[\{\sigma \}]$, which 
corresponds to statistically independent copies of the Hubbard-Stratonovich field, this admittedly more complicated
distribution does not exhibit the factorization to blame for the signal-to-noise
problems present in the approach as originally formulated.

\begin{figure}[t]
\includegraphics[width=1.0\columnwidth]{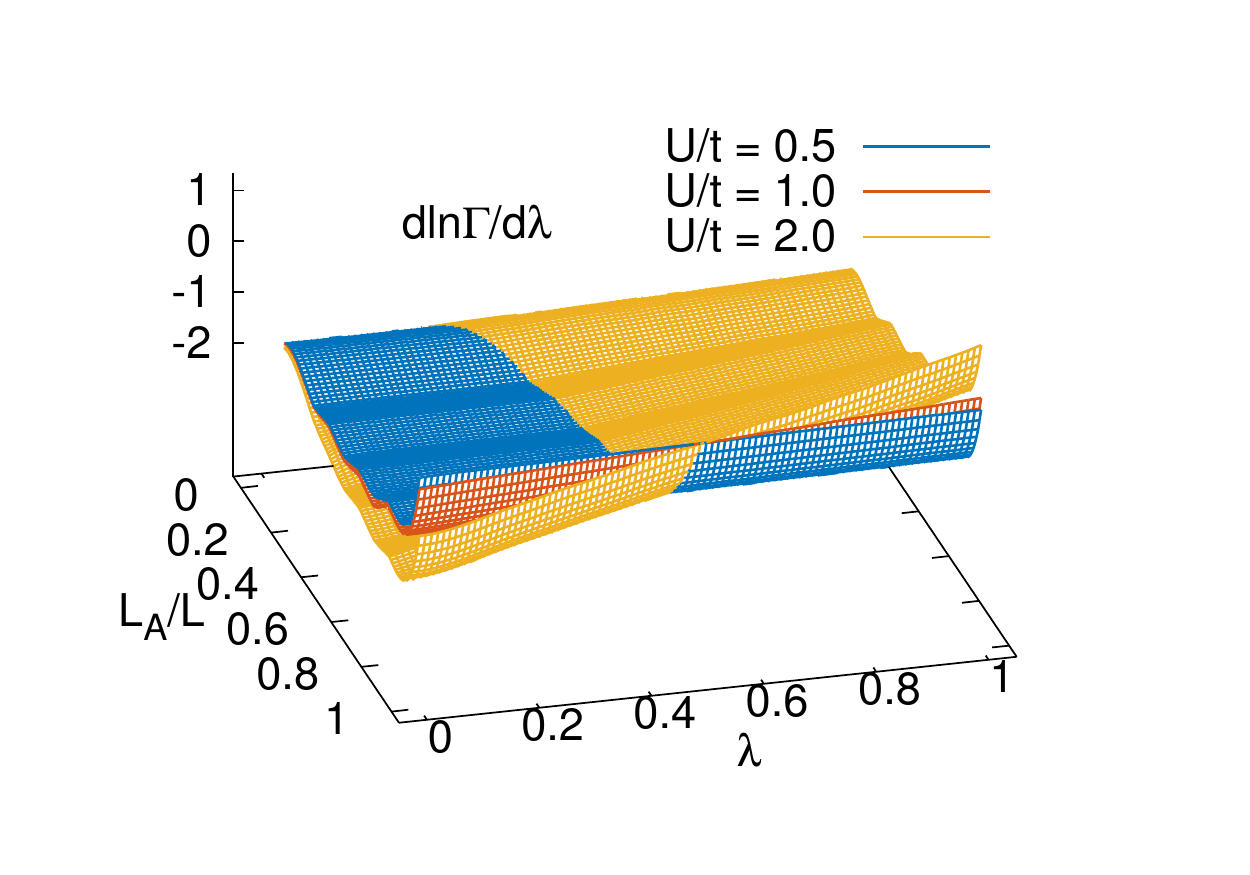}
\caption{\label{Fig:ThreeDCompare}(color online) Stochastic results for $\langle\ln{Q}[\{\sigma \}]\rangle_\lambda^{}$ with $n=2$ for couplings $U/t = 0.5,1.0,$ and $2.0$ as functions of auxiliary parameter $\lambda$ and region size $L_A^{}/L$ for a ten-site Hubbard model.}
\end{figure}

Using Eq.~(\ref{Eq:SnMCFinal}) requires Monte Carlo methods to evaluate 
$\langle\ln{Q}[\{\sigma \}]\rangle_\lambda^{}$ as a function of $\lambda$, followed by integration over $\lambda$. 
As in our previous method, we find here that $\langle\ln{Q}[\{\sigma \}]\rangle_\lambda^{}$ is a smooth function of $\lambda$, 
which is essentially linear in the present case.
It is therefore sufficient to perform the numerical integration using a uniform grid.
The stochastic evaluation of $\langle\ln{Q}[\{\sigma \}]\rangle_\lambda^{}$, for
fixed subregion $A$, can be expected to feature roughly symmetric fluctuations about the mean. As a consequence,
the statistical effects on the entropy are reduced after integrating over $\lambda$.

Finally, we note an interesting application of Jensen's inequality at $\lambda=0$. At that point
\bea
\left . \frac{\partial \ln \Gamma}{\partial \lambda}\right|_{\lambda=0}
&=& \int \mathcal D {\{\sigma\}}^{} P[\{\sigma \}]\; \ln Q[\{\sigma \}] \\
&\leq& \ln \int \mathcal D {\{\sigma\}}^{} P[\{\sigma \}]\; Q[\{\sigma \}] = (1-n)S_n^{} \nonumber,
\eea
which must be satisfied by our calculations. Our Monte Carlo results at $\lambda=0$ indeed satisfy this bound.

\section{Results}

\subsection{Second R\'enyi entropy}

As a first test of our algorithm and in efforts to make contact with previous work~\cite{HMCEE,Grover}, 
we begin by showing results for the second R\'enyi entropy $S_2^{}$ for the one-dimensional Hubbard chain 
with periodic boundary conditions at half filling, whose Hamiltonian is
\beq
\label{Eq:HubbardHam}
\hat H  = - t \sum_{s,\langle ij\rangle}
{\left(\hat c^{\dagger}_{i,s}\hat c^{}_{j,s}+\hat c^{\dagger}_{j,s}\hat c^{}_{i,s}\right)}+
U\sum_{i}{\hat n^{}_{i\uparrow}\hat n^{}_{i\downarrow}},
\eeq
where the first sum includes $s = \uparrow,\downarrow$ and pairs of adjacent sites.
We implemented a symmetric Trotter-Suzuki decomposition of the Boltzmann
weight, with an imaginary-time discretization of $\tau=0.05$ (in lattice units).
As mentioned earlier, the many-body factor in the Trotter-Suzuki approximation was treated
by introducing a replica auxiliary field $\sigma$ for each power of the reduced density matrix. 
As in our previous work, we implemented a Hubbard-Stratonovich transformation of a compact 
continuous form~\cite{MCReviews}.

We present plots for $\langle\ln Q[\{\sigma\}]\rangle_{\lambda}^{}$ with $n=2$ in Fig.~\ref{Fig:ThreeDCompare}. In contrast to the results obtained in Ref.~\cite{HMCEE} and as mentioned above, the resulting expectation demonstrates surprisingly little curvature as the region size $L_{A}^{}$ is varied and is stunningly linear as a function of the auxiliary parameter $\lambda$. Even after twice doubling the strength of the interaction, the curvature of constant-subsystem-size slices is increased only marginally.  We note that if one assumes such benign curvature is a somewhat universal feature, at least for weakly-coupled systems, our method provides a means by which to rapidly estimate the entanglement entropy for a large portion of parameter space at the very least yielding a qualitative picture of its behavior as a function of the physically relevant input parameters.

\begin{figure}[b]
\includegraphics[width=1.0\columnwidth]{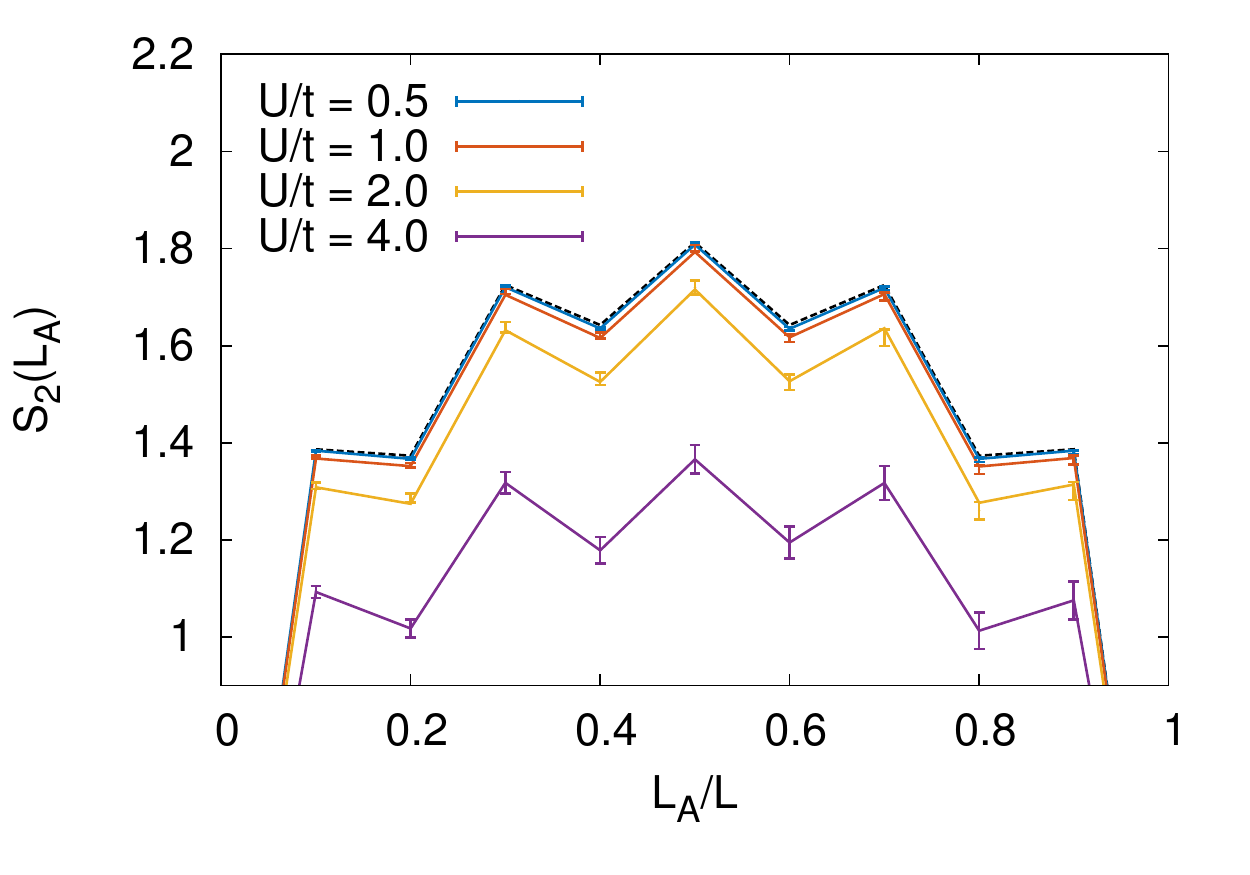}
\caption{\label{Fig:GroverCompare}(color online) Results for the ten-site Hubbard chain for couplings $U/t = 0.5,1.0,2.0$, and $4.0$ for 7,500 samples with associated numerical uncertainties.  Results for $U/t = 0$ are included as a dashed line (black). For all but the largest coupling, exact diagonalization results from Ref.~\cite{Grover} are indicated by solid lines, while for the largest coupling, we provide a line joining the central values of our result to emphasize that its shape is consistent with results for the former.}
\end{figure}

We observe that this surface displays almost no torsion, its dominant features being those present in the noninteracting case i.e. an alternating shell-like structure.  Toward larger region sizes, we observe a combination of twisting and translation culminating in the required, and somewhat delicate, cancellation upon reaching the full system size.  Presented with this relatively forgiving geometry, we performed the required integration via cubic-spline interpolation.  Using a uniformly spaced lattice of size $N_{\lambda}^{} = 20$ points, we determine the desired entropy to a precision limited by statistical rather than systematic considerations.

\subsection{Comparison to exact diagonalization}

\begin{figure}[t]
\includegraphics[width=1.0\columnwidth]{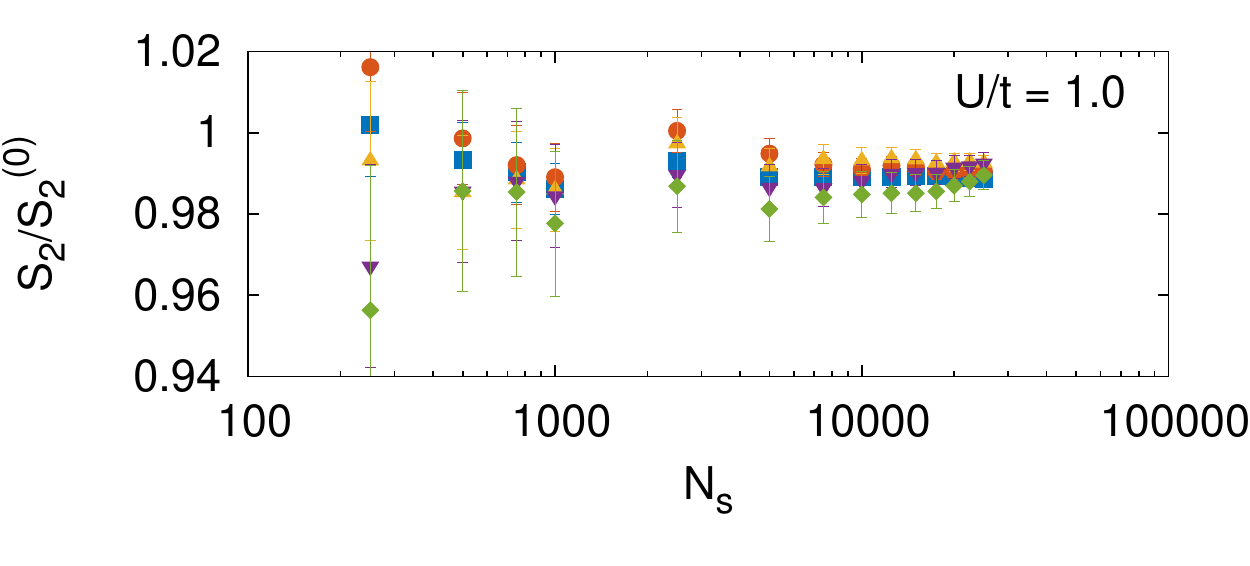}
\includegraphics[width=1.0\columnwidth]{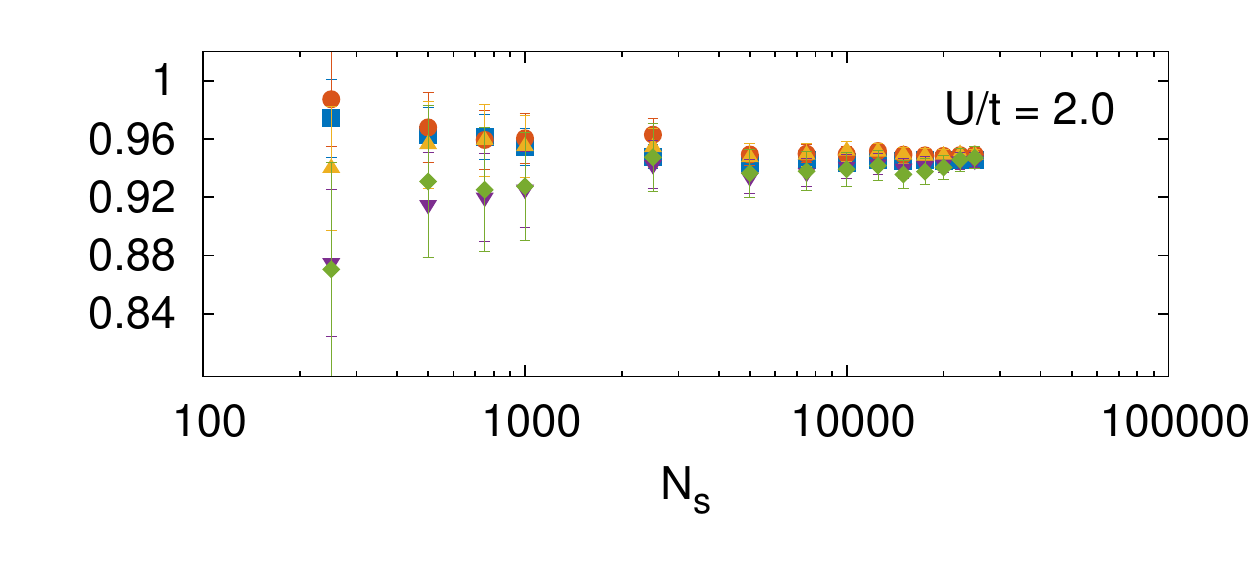}
\includegraphics[width=1.0\columnwidth]{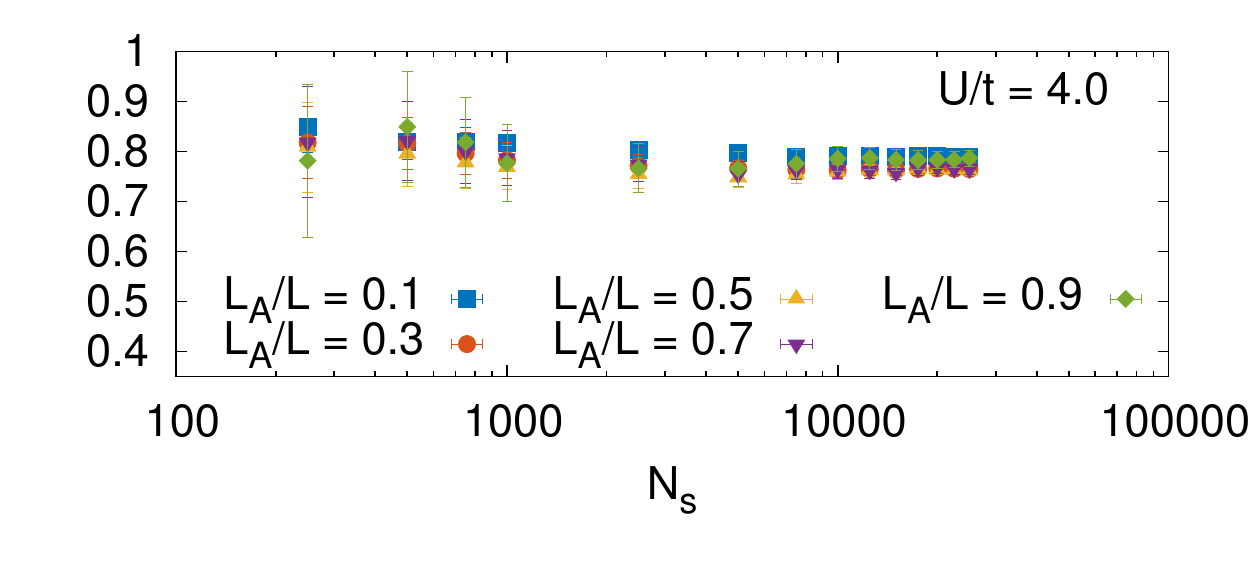}
\caption{\label{Fig:VsSamp}(color online) Entanglement entropy $S_{2}^{}$ in units of the result for a free system plotted as a function of the number of samples $N_{s}$ for couplings $U/t = 0.5,1.0,2.0,$ and $4.0$ demonstrating convergence to within a few percent within the first ten thousand samples.}
\end{figure}

Shown in Fig.~\ref{Fig:GroverCompare} are results for a system of size $L = N_{x}^{}\ell$ with a number of sites $N_{x}^{} = 10$.  For couplings $U/t = 0.5,1.0,2.0$ and $4.0$ and region sizes $L_{A}^{} = 1,2,\dots,10$, we find solid agreement with previous calculations in Refs.~\cite{HMCEE,Grover}, and as in the former, we observe convergence rather quickly with only $O(10^3)$ decorrelated samples as can be seen in Fig.~\ref{Fig:VsSamp}.  Further, for large sample sizes $N_{s}^{}$, we observe that the standard error in the entropy $\Delta S_2^{}$, computed from the envelope defined by the MC uncertainty in the source $\langle\ln Q[\{\sigma\}]\rangle_{\lambda}^{}$ for at each value in $(L_{A}^{},\lambda)$-space, scales asymptotically as $\Delta S_2^{} \thicksim 1/\sqrt{N_{s}^{}}$ up to minute corrections.

%

\subsection{Results for $n\ne2$}

\begin{figure}[t]
\includegraphics[width=1.0\columnwidth]{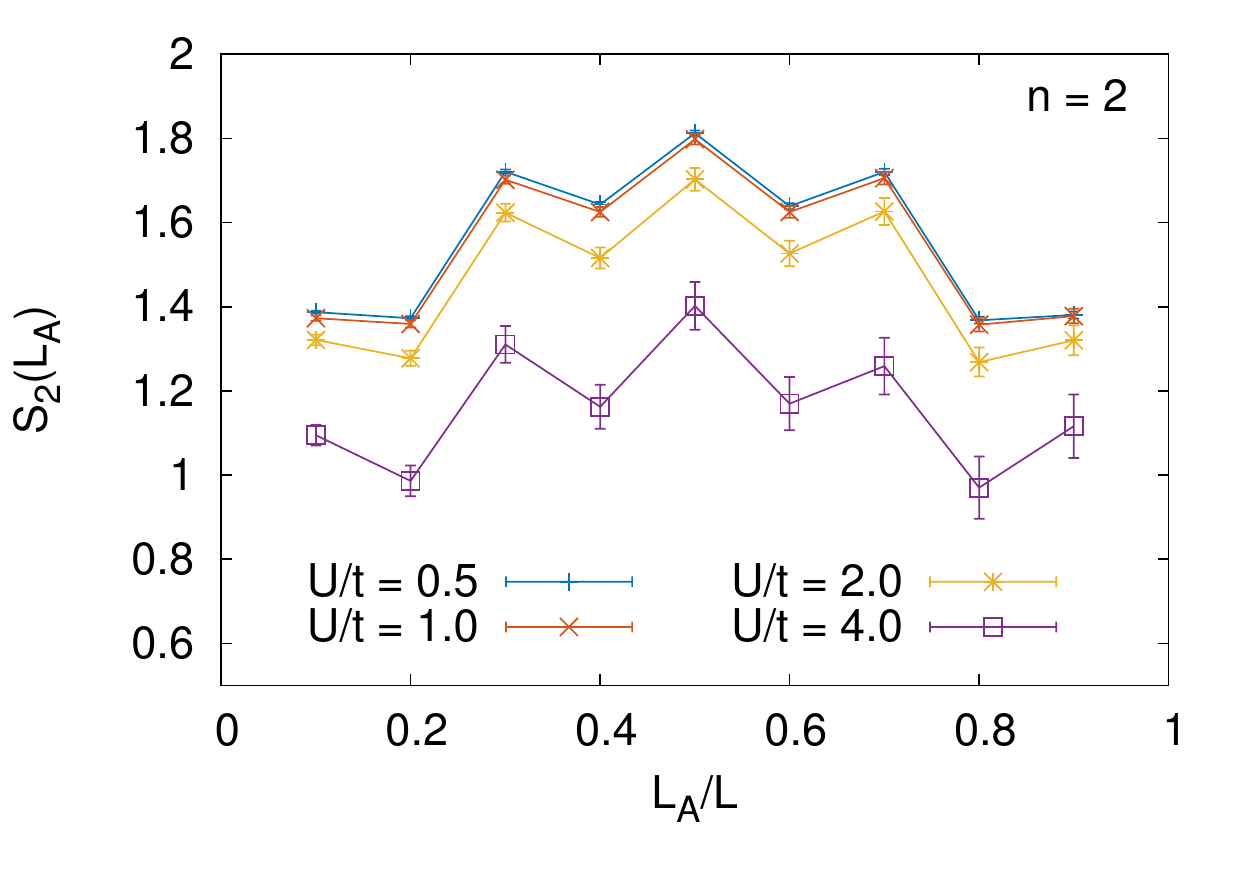}
\includegraphics[width=1.0\columnwidth]{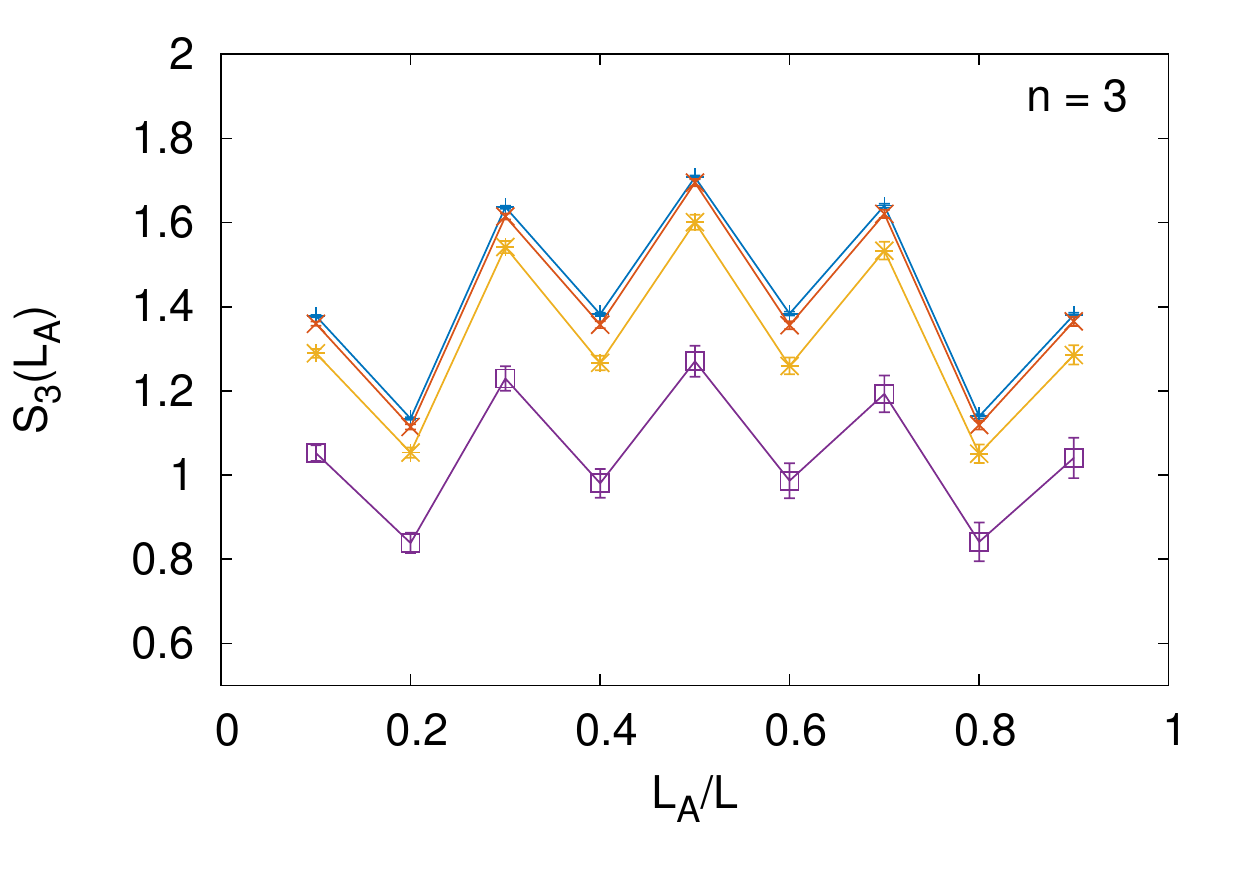}
\includegraphics[width=1.0\columnwidth]{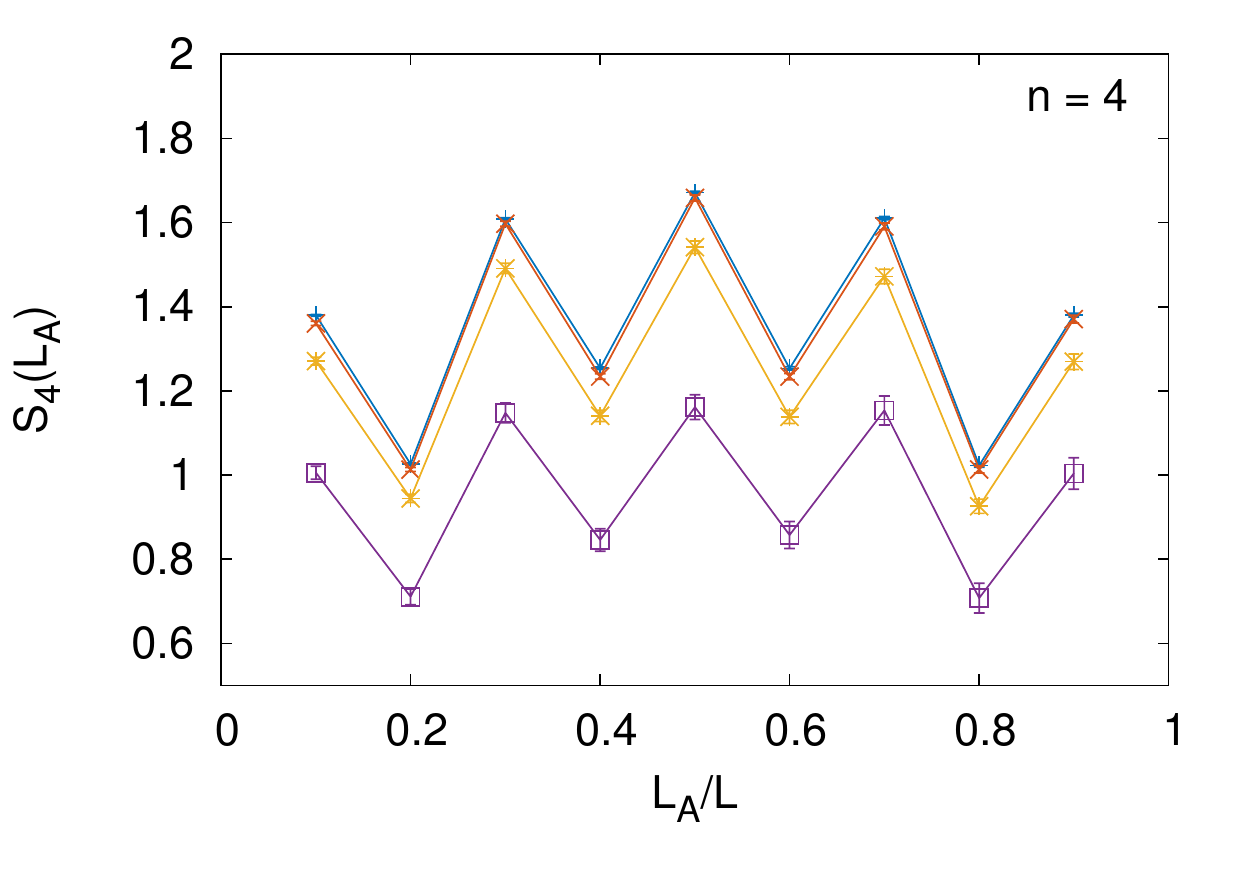}
\caption{\label{Fig:RenyiStack}(color online) R\'enyi entropies $S_n$ for $n=2,3,4$ (top to bottom)
of the 1D attractive Hubbard model, as a function of the subsystem size $L_A/L$. In each plot, 
results are shown for several values of the attractive coupling $U/t$.}
\end{figure}

In this section, we extend the results presented above to $n = 3,4,5,\dots,10$.
In order to highlight the differences between $n=2$ and $n>2$, we
show in Fig.~\ref{Fig:RenyiStack} the R\'enyi entropies $S_n$ for $n=2,3,4$ (top to bottom)
of the 1D attractive Hubbard model, as obtained with our method and the reformulation of the
fermion determinant shown in Eq.~(\ref{MatrixT}).

As evident from the figure, increasing $n$ leads to lower values of $S_n$ at fixed subsystem size $L_A/L$ consistent with knowledge that the R\'enyi entropy is a nonincreasing function of its order.  However, increasing $n$ also amplifies the fluctuations as a function of $L_A/L$.  Interestingly, the approach of our system to the large-$n$ regime is quite rapid, and after only the first few orders, the difference between consecutive entropies is only marginal, most obviously so at weak coupling.  We also observe that, as $n$ is increased, the statistical fluctuations that define the error bars appear to be 
progressively more suppressed, which is particularly evident for the strongest coupling we studied,
namely $U/t = 4.0$.

\begin{figure}[t]
\includegraphics[width=1.0\columnwidth]{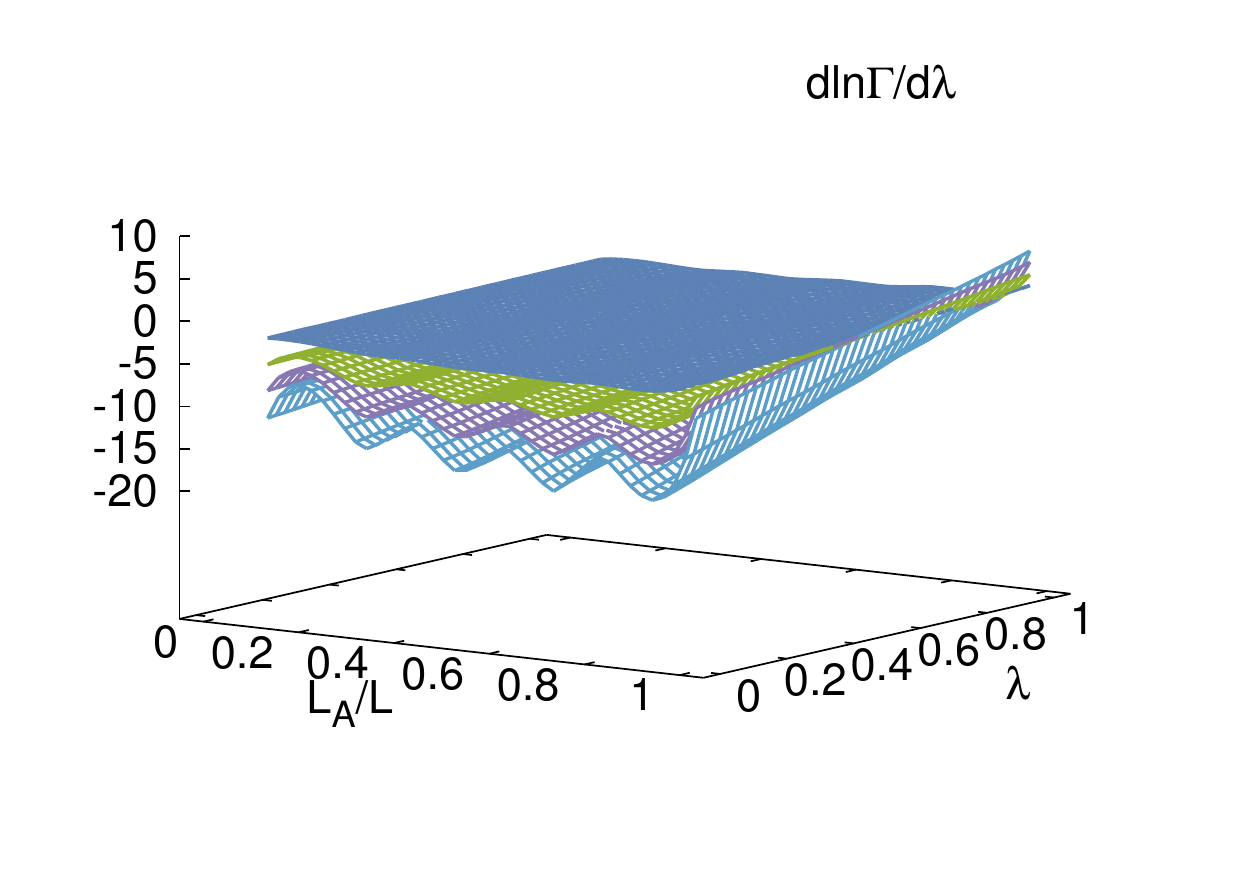}
\caption{\label{Fig:VaryOrderFixCoupling}(color online) Stochastic results for $\langle\ln{Q}[\{\sigma \}]\rangle_\lambda^{}$ with $n=2,4,6,$ and $8$ (top to bottom) for a coupling of $U/t = 2.0$ as functions of auxiliary parameter $\lambda$ and region size $L_A^{}/L$.}
\end{figure}

At the level of the auxiliary function $\langle\ln{Q}[\{\sigma \}]\rangle_\lambda^{}$, we again see very predictable changes in the geometry of this surface as a function both arguments as shown in Fig.~\ref{Fig:VaryOrderFixCoupling}.  With fixed coupling and particle content, increasing the R\'enyi order results in a tilting effect reminiscent of that seen previously with increasing coupling, but rather than being localized away from vanishing subsystem size, the change is much more global, affecting all subsystems in a qualitatively similar fashion and leaving each surface's characteristic quasi-linearity in $\lambda$ intact.  Although the shell-like structure present in this function's $L_A^{}$ dependence is amplified, this increased fluctuation affects the quality of the results negligibly at most, as again, the geometry remains amenable to fairly naive quadratures.

With the data presented above, we would be remiss if we did not attempt an extrapolation not only to the limit of infinite R\'enyi order $S_\infty$, but also to the von Neumann entropy, despite knowledge of the formidable challenges presented by these extrapolations, particularly in the case of the latter.  The former limit provides a lower bound on all finite-order entropies, whereas the latter is of interest to a variety of disciplines and has proven difficult to study.  At fixed coupling and with the knowledge that the R\'enyi entropy is nonincreasing in the order, we found that our results at each fixed region size and at every studied coupling were well-characterized by exponential decays.

\begin{figure}[t]
\includegraphics[width=1.0\columnwidth]{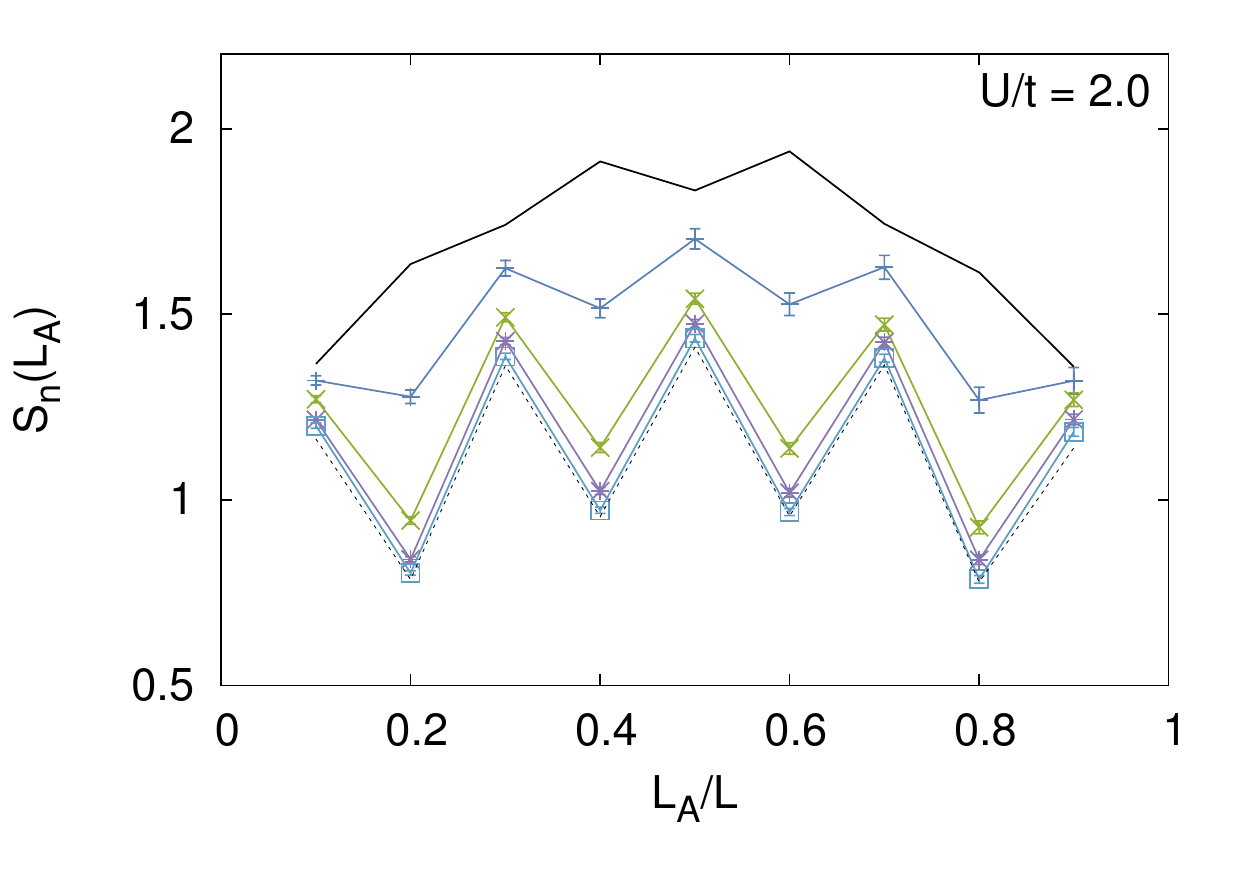}
\caption{\label{Fig:OrderConverge}(color online) 
R\'enyi entropies $S_n$ for $n=2,4,6,$ and $8$ (top to bottom with error bars and colors matching those in Fig.~\ref{Fig:VaryOrderFixCoupling}) of the 1D attractive Hubbard model, as a function of the subsystem size $L_A/L$. 
The solid black line shows extrapolation to $n=1$. The dashed black line shows extrapolation to $n\to\infty$.
Again, results are shown for $U/t=2.0$.}
\end{figure}

\begin{figure}[t]
\includegraphics[width=1.0\columnwidth]{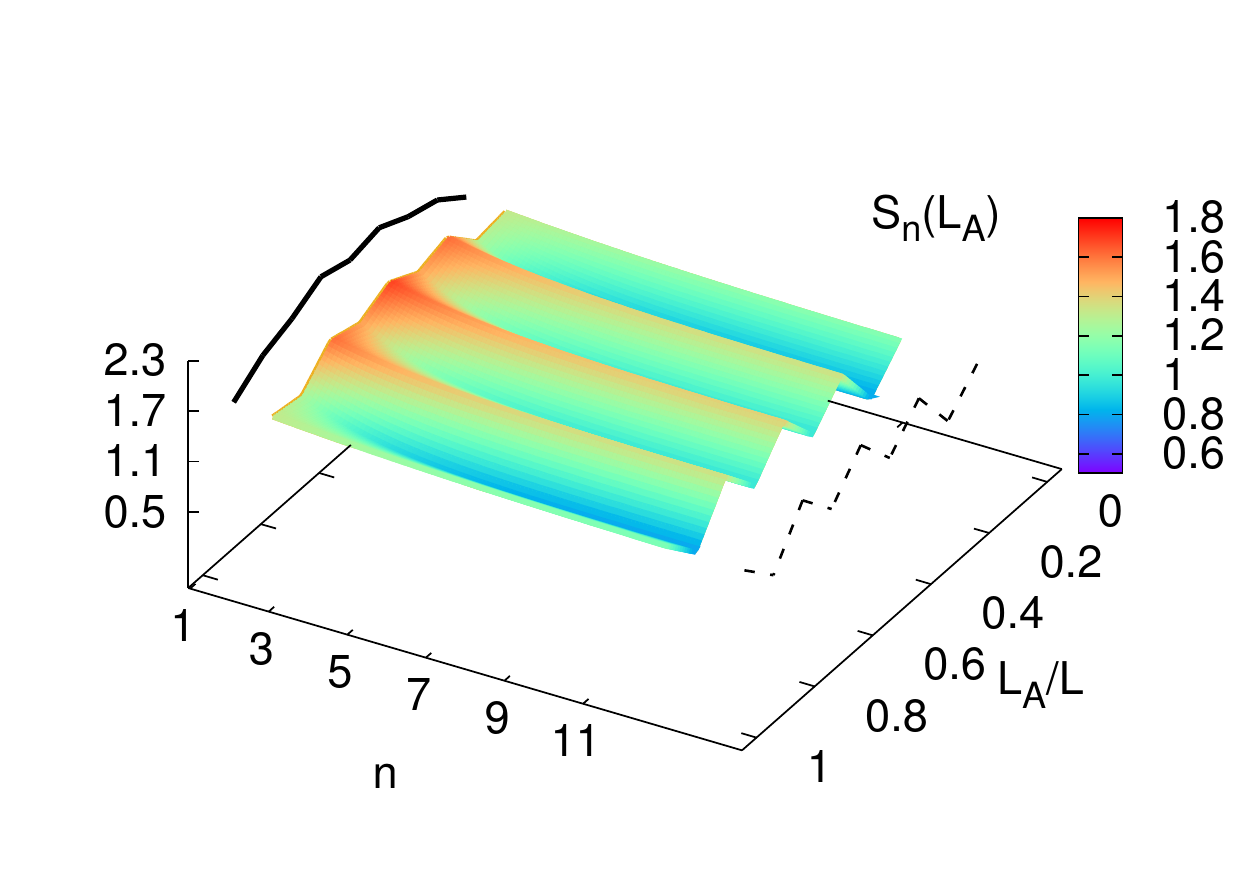}
\caption{\label{Fig:OrderExtrap}(color online) Interpolation of the R\'enyi entropies $S_n$ for $n=2,3,4,\dots,10$ for a coupling of $U/t = 2.0$ given as functions of the auxiliary parameter $\lambda$ as well as the region size $L_A^{}/L$.  An extrapolation to $n=1$ (the von Neumann entropy) as well as to $n\to\infty$ are shown in solid and dashed lines respectively.}
\end{figure}

Interestingly, the relative speed of this decay oscillates as a function of the region size as can be seen in Fig.~\ref{Fig:OrderConverge}.  Regions corresponding to an even number of lattice sites demonstrate a much more sudden initial decay than do those regions comprised of an odd number of sites.  This peculiar oscillation results in an inverted shell structure for the extrapolation to $n=1$, in contrast to the case where $n\to\infty$ in which this feature is preserved.  A representative example of this procedure is shown in Fig.~\ref{Fig:OrderExtrap}.

%

\section{Summary and Conclusions} 

We have presented a method to compute the entanglement entropy of interacting fermions
which takes advantage of an approximate log-normality property of the distribution of
fermion determinants. The resulting approach overcomes the signal-to-noise problem of naive methods, 
and is very close in its core idea to another method we proposed recently: both methods involve defining 
an auxiliary parameter $\lambda$, differentiating, and then integrating to recover $S_n$ after a MC calculation.
The order of the steps is important, as the differentiation with respect to $\lambda$ induces the appearance
of entanglement-sensitive contributions in the MC probability measure.
Beyond those similarities, the present method has the distinct advantages of being simultaneously simpler to formulate
(algebraically as well as computationally) and of explicitly using the approximate log-normality property. Moreover, we 
have found that the $\lambda$ integration step displays clearly more stable numerical behavior in the present approach than
in its predecessor: it is approximately linear in the present case and markedly not so in the original incarnation. 
We therefore strongly advocate using the present algorithm over the former.

In addition to presenting an improved method, we have put forward a straightforward algebraic reformulation of the
equations which, while exactly equivalent to the original formalism, avoids the numerical burden of computing 
inverses of restricted Green's functions in the calculation of $n$-th order R\'enyi entropies for $n>2$. This issue had 
been pointed out by us and others (see e.g. Ref.~\cite{Assaad}) as an inconvenience, as it is perfectly possible
for those matrices to be singular.

As a test of our algorithm, we have presented results for the R\'enyi entropy $S_n$ of the half-filled 1D Hubbard 
model with periodic boundary conditions. The present and old formalisms were used for calculations at $n=2$, which
matched exactly. The rewritten form based on Eq.~(\ref{MatrixT}) was then used to extend our computations to 
$n=3,4,\dots,10$, allowing us to attempt extrapolations in the R\'enyi order in both directions.

Our results show that, with increasing R\'enyi order $n$, the value of $S_n$ decreases for all $L_A/L$, and the fluctuations 
as a function of $L_A/L$ become more pronounced. Remarkably, the statistical MC fluctuations decrease as $n$
is increased. Since the problem we set out to solve was in fact statistical in nature, our observations indicate that 
calculations for large systems and in higher dimensions will benefit from pursuing orders $n > 2$.

\acknowledgements

This material is based upon work supported by the National Science Foundation under
Grants
No. PHY1306520 (Nuclear Theory program)
and
No. PHY1452635 (Computational Physics program). 




\begin{thebibliography}{99}

\bibitem{HMCEE}
J. E. Drut and W. J. Porter,
Phys. Rev. B {\bf 92}, 125126 (2015).


\bibitem{Buividovich}
P. V. Buividovich, M. I. Polikarpov,
Nucl. Phys. B {\bf 802}, 458 (2008).


\bibitem{Melko}
R. G. Melko, A. B. Kallin, and M. B. Hastings, 
Phys. Rev. B {\bf 82}, 100409 (2010);
M. B. Hastings, I Gonz\'alez, A. B. Kallin, and R. G. Melko,
Phys. Rev. Lett. {\bf 104}, 157201 (2010);
S. V. Isakov, M. B. Hastings, and R. G. Melko, 
Nature Phys. {\bf 7}, 772 (2011);
R. R. P. Singh, M. B. Hastings, A. B. Kallin, and R. G. Melko,
Phys. Rev. Lett. {\bf 106}, 135701 (2011);
S. Inglis and R. G. Melko,
Phys. Rev. E {\bf 87}, 013306 (2013).


\bibitem{Humeniuk}
S. Humeniuk and T. Roscilde,
Phys. Rev. B {\bf 86}, 235116 (2012).

\bibitem{McMinis}
J. McMinis and N. M. Tubman,
Phys. Rev. B {\bf 87}, 081108(R) (2013).

\bibitem{Broecker}
P. Broecker and S. Trebst,
J. Stat. Mech. (2014) P08015.

\bibitem{WangTroyer}
L. Wang, M. Troyer,
Phys. Rev. Lett. {\bf 113}, 110401 (2014).

\bibitem{Luitz}
D. J. Luitz, X. Plat, N. Laflorencie, and F. Alet,
Phys. Rev. B {\bf 90}, 125105 (2014).



\bibitem{HMC}
S.~Duane, A.~D.~Kennedy, B.~J.~Pendleton, D.~Roweth,
Phys.\ Lett.\ B {\bf 195}, 216 (1987);
S.~A.~Gottlieb, W.~Liu, D.~Toussaint, R.~L.~Renken,
Phys.\ Rev.\ D {\bf 35}, 2531 (1987).


\bibitem{Grover}
T. Grover,
Phys. Rev. Lett. {\bf 111}, 130402 (2013).


\bibitem{CalabreseCardy}
P. Calabrese, J. L. Cardy, 
J. Stat. Mech. 0406 (2004) P06002.


\bibitem{MCReviews}
D.~Lee, Phys.\ Rev.\ C {\bf 78}, 024001 (2008);
D.~Lee, Prog.\ Part.\ Nucl.\ Phys. {\bf 63}, 117 (2009);
F. F. Assaad and H. G. Evertz, 
Worldline and Determinantal Quantum Monte Carlo Methods for Spins, Phonons and
Electrons, in 
{\it Computational Many-Particle Physics}, H. Fehske, R. Shnieider, and A. Weise
Eds., Springer, Berlin (2008);
J. E. Drut and A. N. Nicholson,
J. Phys. G: Nucl. Part. Phys. {\bf 40}, 043101 (2013);


\bibitem{Peschel}
M.-C. Chung and I. Peschel, 
Phys. Rev. B {\bf 64}, 064412 (2001);
S.-A. Cheong and C. L. Henley, 
Phys. Rev. B {\bf 69}, 075111 (2004);
I. Peschel, J. Phys. A {\bf 36}, L205 (2003).


\bibitem{Assaad}
F. F. Assaad, T. C. Lang, and F. P. Toldin,
Phys. Rev. B {\bf 89}, 125121 (2014);
F. F. Assaad,
Phys. Rev. B {\bf 91}, 125146 (2015).


\bibitem{NoiseSignProblemStatistics}
M. G. Endres, D. B. Kaplan, J.-W. Lee, A. N. Nicholson,
Phys. Rev. Lett. {\bf 107}, 201601 (2011).

\bibitem{LogNormalDeGrand}
T. DeGrand,
Phys. Rev. D {\bf 86}, 014512 (2012).

\end{thebibliography}
\end{document}